\documentclass[%
 amsmath,amssymb,
preprint,
]{revtex4-1}

\usepackage{graphicx}
\usepackage[printfigures]{figcaps}
\usepackage{dcolumn}
\usepackage{bm}

\usepackage[utf8]{inputenc}
\usepackage[T1]{fontenc}
\usepackage{mathptmx}

\usepackage{multirow}
\usepackage{subfig}
\usepackage{color}

\begin{document}

\preprint{The following article has been submitted to The Journal of Chemical Physics }

\title{A comparative first-principles investigation on the defect chemistry of TiO$_2$ anatase.}

\author{Marco Arrigoni  }
 \email{marco.arrigoni@tuwien.ac.at}
\author{Georg K. H. Madsen}%
\affiliation{ 
Institute of Materials Chemistry, TU Wien, A-1060 Vienna, Austria
}%

\date{\today}

\begin{abstract}
Understanding native point defects is fundamental in order to comprehend the properties of TiO$_2$ anatase  in technological applications. Several first-principles studies have been performed in order to investigate the defect chemistry of this material. The reported values are, however, scattered over a wide range.  In this manuscript we perform a comparative study employing different approaches based on semilocal, DFT+$U$ and screened hybrid functionals in order to investigate the dependence of defect properties, such as formation energies and charge transition levels, on the employed computational method. While the defects in anatase, like in most transition-metal oxides, generally induce the localization of electrons or holes on atomic sites, we notice that, provided an alignment of the valence bands has been performed, the calculated defect formation energies and transition levels using semi-local functionals are in a fair  agreement with those obtained using hybrid functionals. A similar conclusion can be reached for the thermochemistry of the Ti-O system and the limit values of the elemental chemical potentials. We interpret this as a cancellation of error between the self-interaction error and the overbinding of the O$_2$ molecule in semi-local functionals.
Inclusion of a $U$ term in the electron Hamiltonian offers a convenient way for obtaining more precise geometric and electronic configurations of the defective systems.
\end{abstract}

\maketitle

\section{ \label{intro} Introduction}
TiO$_2$ is one of the most representative materials in photocatalysis \cite{Linsebigler-1995,Schneider-2014}, with the anatase polymorph being the most commonly found  in synthesized nanoparticles where it shows a higher photocatalytic activity than the  thermodynamical stable rutile bulk phase \cite{Sumita-2002,Luttrell-2014}. Anatase is also attracting interest as a promising and inexpensive  transparent conducting oxide due to its wide band gap (around 3.2 eV \cite{Kavan-1996}) and large intrinsic $n$-type carrier concentrations \cite{Forro-1994}. The conductivity decreases after annealing in ambient atmosphere at high temperature \cite{Forro-1994}, but can be further enhanced through doping with group-V elements \cite{Furubayashi-2005,Hitosugi-2005}. TiO$_2$ is often produced in a reduced state, where it shows states in the gap and a pale-blue color \cite{Diebold-2003}. The non-stoichiometry of reduced TiO$_2$ has been associated to both oxygen vacancies, titanium interstitials and surface hydroxyl groups \cite{Wendt-2008, Yin-2018}.  

In order to explain the observed features, several first-principles studies have considered the various intrinsic defects of anatase. They tend to agree on the fact that the most relevant electron donors, which govern the $n$-type behaviour of intrinsic anatase, are oxygen vacancies ($\square_\mathrm{O}$) and titanium interstitials (Ti$_i$) and the most important electron acceptor are the titanium vacancies ($\square_\mathrm{Ti}$).  However, one finds that the values of  important quantities, such as defect formation energies and thermodynamic charge transition levels, are very scattered. To emphasize this, we mention how the description of Ti$_i$ and $\square_\mathrm{O}$ defects may be different both  quantitatively and qualitatively.
 Na-Phattalung \emph{et al.}  employed the local density approximation (LDA) and predicted that both Ti$_i$ and $\square_\mathrm{O}$ are very shallow donors\cite{Na-2006}. On the contrary Osorio-Guill{\'e}n \emph{et al.}, using a generalized gradient approximation (GGA) functional, found that Ti$_i$ and $\square_\mathrm{O}$ behave like deep donors with transition levels more than 1~eV below the conduction band minimum (CBM) \cite{Osorio-2008}. Morgan and Watson used the GGA+$U$ method and also found Ti$_i$ and $\square_\mathrm{O}$ to be deep donors \cite{Morgan-2009}. Most recently, Boonchum \emph{et al.}\cite{Boonchun-2016} and Deak \emph{et al.}\cite{Deak-2015} used the same screened hybrid functional (HSE06 \cite{Heyd-2003}) to study Ti$_i$ and $\square_\mathrm{O}$. While Boonchum \emph{et al.} found $\square_\mathrm{O}$ to be a very shallow donor and Ti$_i$ to have transition levels located no further than 0.4 eV below the CBM, than  Deak \emph{et al.} found  transitions between +2/+1/0 charge states of $\square_\mathrm{O}$ to be  0.4~eV and 0.05 eV below the CBM and +4/+3/+2/0 transition levels for Ti$_i$ to be in the range between 1.3 eV and 0.3 eV below the CBM\cite{Deak-2015}.


Such discrepancies are problematic if one aims to predict the properties of technologically relevant materials by means of first-principles simulations. They are, however, not completely unexpected as the above-mentioned studies employ very different computational methods, involving not only different exchange-correlation (xc) functionals, but also different correction schemes for charged defects, supercell sizes and pseudopotentials.  

Regarding the choice of the xc functional, it is well known that common local and semi-local functionals suffer from two main shortcomings affecting the  description of point defects in semiconductor materials \cite{Lany-2008,Deak-2011}. The self-interaction error, due to which portions of the charge density associated with a given electron tend to repel themselves, yields  electronic configurations with an exaggerated electron delocalization for those defects which induce localized defective states.
Furthermore, the underestimation of the fundamental gap results in a tendency to mix the defect-induced states with the band edges, yielding an erroneous delocalization of the defect-induced charge density in valence-band-like or conduction-band-like states. 

Both these shortcomings can be traced to a derivative discontinuity of the exact Kohn-Sham xc functional at integer particle numbers which cannot be reproduced within the LDA and GGAs\cite{Perdew_PRL82}. To ameliorate them, different approaches have been proposed. Nowadays, hybrid functionals are considered as the method of choice for the first-principle study of point defects in solids since the incorporation the exact exchange partly corrects the self-interaction and introduces an approximated derivative discontinuity for the xc energy. As a consequence, localized states can be described more accurately and the predicted band gap is much closer to the experimental value. In particular, screened hybrid functionals are currently preferred for solids due to their superior accuracy and reduced computational costs with respect to non-screened hybrids \cite{Janesko-2009}.

While hybrid functionals seem a suitable choice for the description of point defects, the computational costs involved are still high. The problem is particularly significant for the study of defects, since these require relatively large supercells and non-trivial structure relaxations that can be difficult to explore systematically. A good compromise between computational costs and reliability might be offered by the DFT+$U$ approach, which aims at a correct description of the derivative discontinuity by the introduction of an on-site $U$ term for localized electrons\cite{ldaUfll}.

Calculated defect energies are influenced by other factors than the xc functional. As mentioned already, even the two studies employing the same hybrid functional reached different conclusions \cite{Boonchun-2016,Deak-2015}. 
It would thus be important to isolate the role of the chosen xc functional, the most important approximation present in the Kohn-Sham scheme, from other computational parameters in predicting the properties of point defects in an technologically relevant materials such as TiO$_2$ anatase.
 Recently, it has been shown, for certain materials and localized defects, that a good agreement between the the thermodynamic charge transition levels predicted by semi-local and hybrids functionals can be obtained if the values of the electron chemical potentials are given with respect to a common reference \cite{Alkauskas-2008,Alkauskas-2011,Ramprasad-2012,Freysoldt-2016,Lyons-2017}.

In the present study, we thus investigate the defect chemistry of intrinsic anatase employing different functionals at the semi-local, DFT+$U$ and hybrid level. By comparing the results, we  estimate the level of agreement between different xc functionals. We find that altough PBE+$U$ gives an electronic and geometric structure in better agreement with HSE15, predicted charge transition levels and defect formation energies are in better agreement between semilocal and HSE15 functionals. It thus appears that while PBE+$U$ would be very useful to estimate the geometric and electronic structure of the defect, which can be used as a starting point for more accurate theoretical approaches, standard GGA functionals are more suitable for a first estimation of the energetic properties of the defect. 

\section{ \label{method} Computational Method}

\subsection{\label{sec:dilute} Defect Formation in the Dilute Limit}
For the formation of  point defects the most appropriate thermodynamic potential is the grand potential \cite{Zhang-1991}. Therefore it is natural to define the defect energy, $\Delta E_d(D^{(q)})$, of a given point defect $D$ in the charge state $q$ as the change in grand potential after  the introduction of the defect in the pristine host material:
\begin{equation}
\label{eq:formene}
\Delta E_d(D^{(q)}) = \Delta E_f(D^{(q)}) - \sum_i n_i \Delta\mu_i + q \mu_e.  
\end{equation}
where $\Delta E_f(D^{(q)})=E(D^{(q)})-E_\mathrm{bulk}-\sum_i n_i E_i$ is the defect formation
energy with respect to the reference states of the parent elements. $E(D^{(q)})$ is the free energy of the supercell containing the point defect, $E_\mathrm{bulk}$ is the free energy of the supercell describing the pristine material, $n_i$ is the number of atoms of type $i$ which need to be removed ($n_i < 0$) or added ($n_i > 0$) to the system in order to create the point defect and $E_i$ the energy of the standard state. $\Delta \mu_i$ is the change in chemical potential of the element $i$ from the standard state and $\mu_e$ is the chemical potential of the electron.
As a common approximation, we replace the Gibbs free energy of the solid with the ground-state electronic energy computed by first-principles. Even tough both harmonic\cite{Arrigoni-2015,Bjor-2016,Arrigoni-2016} and anharmonic\cite{Glensk-2014} contributions can have a non-negligible effect on the bulk energies, these are mainly relevant at high temperatures. 

The $E(D^{(q)})$ must be corrected for the finite-size-errors which arise in the supercell method \cite{PDRev}. While most of the errors can be minimized by using a large enough supercell, in the presence of charged defects, electrostatic interactions are too long-ranged and cannot be neglected for any realistic supercell size. Among the various methods proposed in the literature for correcting the electrostatic finite-size effects, we employ the one proposed by Kumagai and Oba, which has proved to be very effective for anisotropic systems and for ionic materials, where large atomic relaxations induced by the presence of a point-defect make the use of other potential alignment methods difficult \cite{Kumagai-2014}.

A thermodynamic charge transition level is defined as the value that the electron chemical potential must have in order for two different charge states, $q$ and $q'$, of a defect to have the same defect energy. It is customarily to express the electron chemical potential in terms of the valence band maximum of the host material, $\epsilon_V$, and the Fermi level, $E_F$, which varies between zero and the band gap of the material: $\mu_e = \epsilon_V + E_F$. With this convention and using equation~\eqref{eq:formene}, we can write  the charge transition levels as:
\begin{equation}
\label{eq:translevel}
\epsilon_0(q/q') = \frac{E(D^{(q)}) - E(D^{(q')})}{q' - q}- \epsilon_V ,
\end{equation}
 This expression emphasizes how the charge transition levels do not depend on the chemical potentials of the elements but only on the valence band maximum eigenvalue.

\subsection{\label{sec:chempots} Chemical Potentials}
The values of $\Delta \mu_i$ are important, not only because they enter in equations \eqref{eq:formene}, but also because they give a connection between the first-principles defect calculations and the experimental growth conditions of the system.

Thermodynamic equilibrium constraints the possible values the chemical potentials of the elements can assume. For TiO$_2$ anatase the constraints are given by the following conditions:
\begin{subequations}
\label{eq:atchempots}
\begin{gather}
\label{subeq:equality}
\Delta \mu_\text{Ti} + 2 \Delta \mu_\text{O} = \Delta h_f(\mathrm{TiO}_2; \mathrm{anatase}),\\
\label{subeq:inequality}
x \Delta \mu_\text{Ti} + y \Delta \mu_\text{O} \leq  \Delta h_f(\mathrm{Ti}_x\mathrm{O}_y),\\
\label{subeq:bounds}
\Delta \mu_\text{Ti} \leq 0, \quad \quad
\Delta \mu_\text{O}  \leq 0
\end{gather}
\end{subequations}
For Ti we took as standard state the HCP titanium structure and for  O we took as the gas phase of the O$_2$ molecule in its triplet ground state.  
$\Delta h_f$ is the enthalpy of formation (per formula unit) of the compound of interest. 
Equation \eqref{subeq:equality} represents the thermodynamic stability condition for TiO$_2$ anatase and shows that only one of the elemental chemical potential is an independent variable. 
Equations~\eqref{subeq:inequality} and \eqref{subeq:bounds} state the fact that we consider only those thermodynamical states where anatase is stable. Note that while at standard pressure and temperature the thermodynamical stable phase of TiO$_2$ is rutile, this is not the case within the GGA, GGA+$U$ and hybrid functionals approximations, as it is shown in section~\ref{sec:res_chempots}. 
An accurate evaluation of the formation enthalpies of the various oxides is critical for the determination of the chemical potential of Ti. We will discuss these points for the Ti-O system in section \ref{sec:res_chempots}.

\subsection{\label{sec:compdetails} Computational Details}
We performed the first-principles calculations employing different xc functionals. A series of calculations was done employing the version of the PBE functional revised for solids (PBEsol) \cite{Perdew-2008} while other two series employed the DFT+$U$ formalism in its rotationally invariant fully localized limit\cite{ldaUfll,Dudarev-1998}. Finally, for the most relevant electron donor and acceptors, \emph{i.e.} Ti$_i$ and $\square_\mathrm{O}$ and $\square_\mathrm{Ti}$, respectively, we also employed the HSE functional. Since the standard HF admixture of 25\% overestimates the band gap by around 0.5 eV, we employed a value of 15\% which predicts a band gap of 3.12 eV, in good agreement with the experimental value of around 3.2 eV.  In the text, such a parametrization of the HSE functional is denoted as HSE15.

The first batch of DFT+$U$ series employs the PBE functional and a value of $U$ equal to 5.8 eV which is applied on the $3d$ states of all Ti atoms of the titanium oxides (except on TiO which is metallic). We denote such approach as PBE+$U$[Ti]. The value of $U$ was chosen because it offers a compromise between accurately described cell parameters and the band gap (see Table \ref{table:gs}). This value is also the value found for $d$ orbitals in TiO from constrained DFT calculations \cite{Anisimov-1991}. The second batch of DFT+$U$ calculations also adds a $U$ term on the $2p$ O orbitals for the same oxides (PBE+$U$[Ti,O]). This was proposed by Morgan and Watson  to be necessary in order to correctly describe ionized acceptors such as $\square_\mathrm{Ti}$\cite{Morgan-2009}. For the oxygen $2p$ orbitals we use the same value of $U$ of 5.25 eV proposed by these authors.  

All calculations were performed using the projector augmented-wave method \cite{Bloch-1994} as implemented in the computer code VASP \cite{Kresse-1996}. As valence electrons, we considered the $3p$, $3d$ and $4s$ ones for Ti and the $2s$ and $2p$ ones for O. 
Plane waves up to an energy cutoff of 500 eV were included in the basis set. The calculations on the conventional cell of TiO$_2$ anatase employed a $6 \times 6 \times 2 $ $\Gamma$-centered grid for reciprocal space integration.  We checked the convergence of the internal energy up to a large $16 \times 16 \times 8 $ $\Gamma$-centered grid and found a difference of less than 2 meV per atom. Since the valence band maximum of anatase does not lie at any high-symmetry point in reciprocal space, to obtain an accurate estimation of this eigenvalue we calculated the band structure of the primitive cell. For calculations involving other phases in the Ti-O system, we used a $\Gamma$-centered grid with at least 1000/$n$ $k$-points, where $n$ is the number of atoms in the simulation cell. Ionic positions and cell parameters of the pristine systems were optimized until forces on all atoms were below 0.01 eV/\AA. The convergence threshold for the electronic energy was set to 10$^{-5}$ eV. Spin polarization was allowed in all calculations.

The dielectric tensor, both ionic and electronic contributions,  necessary in order to calculate $E_{corr}$ of equation \eqref{eq:formene}, were calculated using density functional perturbation theory \cite{Baroni-1986,Gadjo-2006} using a dense $24 \times 24 \times 8$ $\Gamma$-centered $k$-point grid. For HSE15 the experimental value has been used.

Point defects were modeled employing a $3 \times 3 \times 1 $ expansion of the conventional tetragonal anatase cell. Such supercells contain 108 atoms. The number of $k$ points for reciprocal space integration was reduced accordingly, except for the HSE15 calculations, where due to high computational costs, we employed a $\Gamma$-only grid. Ionic positions were optimized using a conjugated-gradient method keeping the same thresholds for forces and total energies as for the pristine systems. The cell parameters were kept fixed to the ground-state values obtained for the pristine system using the corresponding xc functional.

A quite used practice consists in fixing the cell parameters of defective supercells to experimental values or to values obtained from different functionals. This practice is  quite common, for example, when as a starting point for more expensive approaches (\emph{e.g.} hybrid functionals) one takes the structures optimized with less time consuming functionals. In practice such an approach induces a pressure on the simulation cell, which can be considerable if we consider the fact that different functionals can disagree on the cell parameters by around 2-3\%. To assess the validity of such methods, we calculated the formation energies of point defects in the PBE+$U$[Ti] setup both fixing the cell parameters to the ground-state PBE+$U$[Ti] values and to the PBEsol values, which are very close to the experimental ones.

In order to compare charge transition levels calculated using different xc functionals, we align the top of the valence bands predicted by the different functionals to the vacuum level by calculating ionization potentials (IPs). To perform such an alignment we use the three-step approach proposed in reference \onlinecite{Stefanovic-2014}. 
Such approach requires the calculation of the valence-band-maximum eigenvalue, $\epsilon_V$, in the bulk system  and the calculation of the averaged electrostatic potential in the vacuum region and in a bulk-like region of a sufficient thick slab.  We perform slab calculations for the PBEsol, PBE+$U$[Ti], PBE+$U$[Ti,O] and HSE15 functionals using a 10-atomic-layers slab presenting the non-polar (101) surface of anatase, which is the most stable one \cite{Diebold-2003}. Slabs are separated by their periodic images along the direction perpendicular to the surface by 50 \AA \, of vacuum. Ionic positions were not relaxed as relaxation effects are generally small \cite{Lyons-2009}.  Using slabs, and assuming that the surface charge densities are described in a similar way by different functionals and surface dipoles are small, it is possible to reference $\epsilon_V$ to the vacuum level by comparing the value of the electrostatic potential obtained in a region far from the surface, which represents the vacuum, to the averaged electrostatic potential in the bulk-like region of the slab. Such procedure gives the IP as predicted by a given functional. Once IPs have been calculated for  each xc functional, they can be compared in order to align the obtained valence band maxima  to an unique reference, which we take as $\epsilon_V$ of the hybrid functional, since hybrids give much more reliable IPs than standard semi-local functionals \cite{Hinuma-2014}. The alignment is then given by: $\Delta \epsilon_V = $ IP - IP(Hyb.). Where IP(Hyb.) indicates the ionization potential calculated with the hybrid functional HSE15.

\section{ \label{results} Results}

\subsection{\label{sec:res_bulk} Bulk Properties} 
\begin{table}
\caption{\label{table:gs} Bulk properties of TiO$_2$ anatase. $a$ and $c$ are the cell parameters of the tetragonal cell, $E_g$ is the fundamental gap, $\epsilon^\infty_{\nu \nu}$ and $\epsilon^0_{\nu \nu}$ are the symmetry-independent components  of the (static) electronic and ionic contributions, respectively, of the dielectric tensor. $\Delta \epsilon_V $ is the offset of the valence band maximum with respect to the value calculated for HSE15. A negative sign means that the HSE15 valence band maximum lies deeper than the valence band of the considered functional. }
\begin{ruledtabular}
\begin{tabular}{lrrrrr}
Property & PBEsol & PBE+$U$[Ti] & PBE+$U$[Ti,O] & HSE15 & Exp.\\
\hline
$a$ (\AA) &3.77 & 3.88 & 3.86 & 3.78  & 3.79\footnote{Ref. \onlinecite{Rao-1970} at $\approx$ 300K} \\
$c$ (\AA) & 9.56  &  9.77 &  9.74 & 9.64 &  9.54$^{\mathrm{a}}$ \\
$E_g$ (eV)  & 2.08  &  2.79  & 2.95 & 3.12  & 3.2\footnote{ Ref. \onlinecite{Kavan-1996}} \\
 $\epsilon^\infty_{x x}$& 6.90  &   5.64   & 5.42 & & 5.82\footnote{ Ref. \onlinecite{Gonzalez-1997}}   \\
 $\epsilon^\infty_{z z}$ &  6.32 &  5.48   &  5.30 & &  5.41$^{\mathrm{c}}$  \\
 $\epsilon^0_{x x}$ &  50.88     &    18.98  &  17.68& &  45.1$^{\mathrm{c}}$  \\
 $\epsilon^0_{z z}$&  22.93      &     12.62 &  12.00&  &22.7$^{\mathrm{c}}$ \\
$\Delta \epsilon_V $ (eV) & -0.70 & -0.78 & -0.64 & 0 & 
\end{tabular}
\end{ruledtabular}
\end{table}

TiO$_2$ anatase crystallizes in a body-centered tetragonal unit cell (space group $I4_1/amd$) containing 12 atoms. The top of the valence band is mainly formed by overlapping oxygen $2p$ orbitals, while the bottom of the conduction band mostly by  titanium $3d$ orbitals.
Various properties of bulk anatase, calculated with different functionals, are summarized in Table \ref{table:gs}.  

PBEsol and HSE15 give cell parameters in very good agreement with the experimental values, but  HSE15 slightly overestimates $c$ by 1\%.  PBE+$U$ overestimation is larger: the cell parameter $a$ is overestimated by around the 2.4\% in PBE+$U$[Ti] and the 1.8\% in PBE+$U$[Ti,O]; while the $c$ parameter is overestimated by around 2.4\% and 2.1\%, respectively. The PBEsol band gap is severely underestimated by 35\%, as it is expected for semilocal functionals. Both  PBE+$U$[Ti] and PBE+$U$[Ti,O] give a value closer to the experimental one (smaller than $\approx$ 13\% and 8\%, respectively). The effect of an $U$ term on the Kohn-Sham effective potential is to introduce a repulsive term for less than half-filled states and an attractive one for the other ones\cite{ldaUfll,Himmetoglu-2014}. Therefore, adding $U$ to the empty Ti-$3d$ states pushes the conduction band maximum up, opening the Kohn-Sham gap; adding an $U$ term also on the occupied O-$2p$ states will push the valence band down, opening the gap even more. This fact can also be noticed by considering the last row of Table~\ref{table:gs} which shows that PBE+$U$[Ti,O] valence band maximum lies indeed lower than the PBE+$U$[Ti] one. Note that for this material the standard parametrization of HSE06 overestimates the band gap by around 17\% \cite{Boonchun-2016}. The IP calculated for the (101) surface of anatase, using the HSE15 functional, has a value of 7.67 eV, in close agreement with the value predicted for this surface by taking into account many-body effects at the GW level \cite{Stefanovic-2014}, indicating that  HSE15 offers a good reference for comparing valence band maxima.

Regarding the computed dielectric tensor, the electronic contribution calculated with PBE+$U$[Ti] and PBE+$U$[Ti,O] agrees well with the experimental values; while PBEsol tends to overestimate it by an average of the 18\%. On the other hand PBE+$U$ severely underestimate the ionic contribution, while PBEsol gives values in better agreement with the experiments. The values of the dielectric tensor are needed only for correcting the electrostatic finite-size-effects in supercell calculations in the correction scheme of Kumagai and Oba \cite{Kumagai-2014}. Therefore, the computed value is the correct one to be used, since it describes consistently with the employed xc functional the medium response to the electric field generated by the array of charged defects,  modeled as point charges. 

\subsection{\label{sec:def_chemistry} Defect Chemistry}

\begin{figure}
\includegraphics[width=1\textwidth]{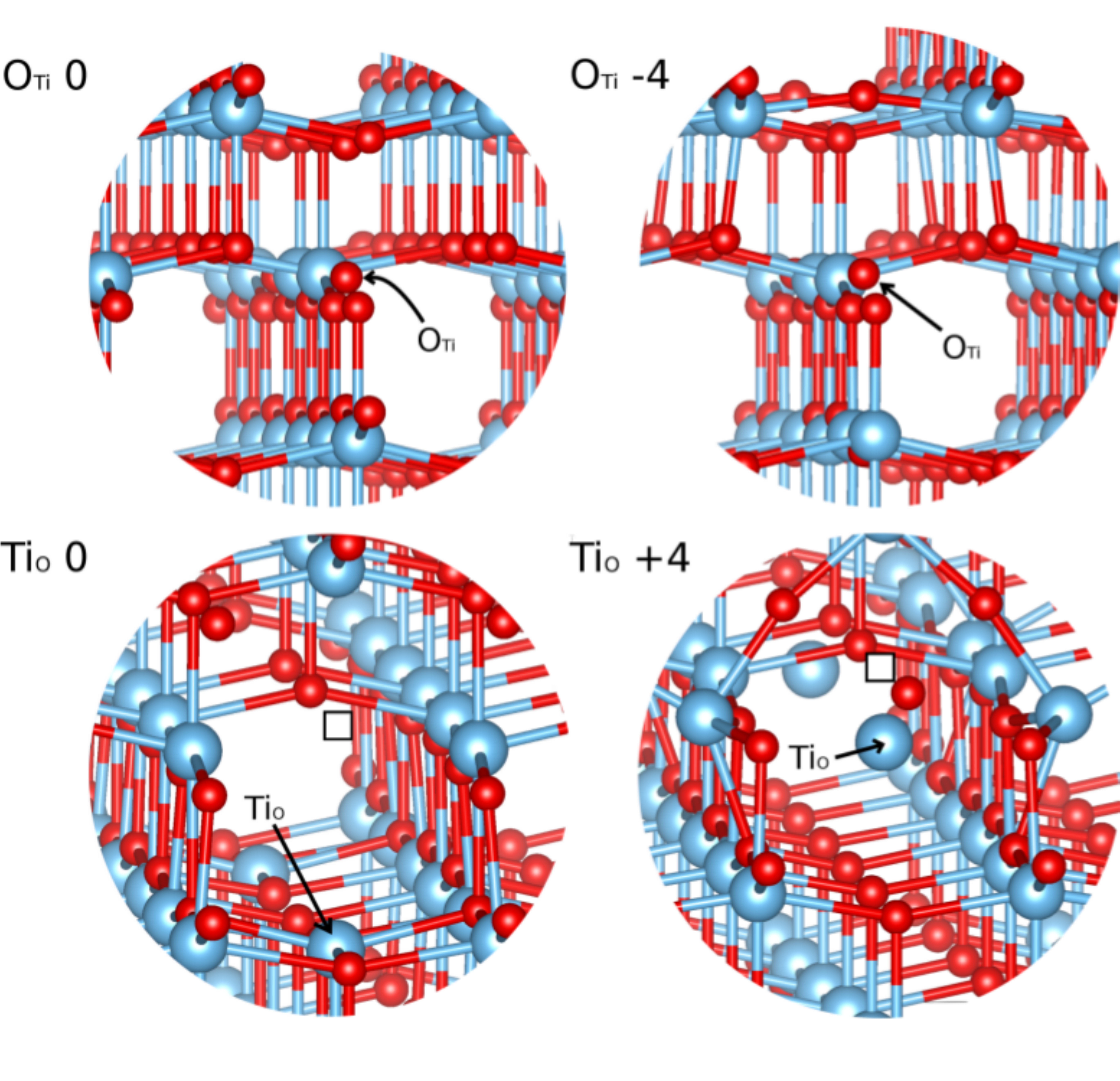}
\caption{\label{fig:antisites} O$_\mathrm{Ti}$ (top) and  Ti$_\mathrm{O}$ (bottom) antisites in the neutral  and fully ionized states. The empty square box represents the oxygen vacancy left by the Ti atom after the Ti$_\mathrm{O}$ defect relaxes. Oxygen atoms are represented in red. The picture was produced using the VESTA software \cite{Momma-2011}.}
\end{figure}

In our comparative study  we take into account a wide range of point defects and charge states and compare the results keeping the same computational parameters, except for the xc functional (and $k$-point mesh for hybrid functional calculations), in order to assess the dependency of the calculated values on the choice of the xc functional itself. We considered the largest number of defects within the PBE+$U$[Ti] scheme. In particular, we studied: Ti$_i$ as an intrinsic donor in charge states 0, +1, +2, +3, +4. The defect was placed in an interstitial site where it obtained a quasi-pyramidal coordination \cite{Finazzi-2009} and occupies the $8e$ Wyckoff site.  $\square_\mathrm{Ti}$ was considered as an acceptor in the charge states 0, -1, -2, -3, -4, its Wyckoff site is $4a$. O$_i$ was studied in the -2, -1, 0, +1, +2 charge states. O$_i$gives rise to a O$_2$ dimer whose bond length depends on the charge state \cite{Kamisaka-2011}. Its center of mass occupies the $8e$ Wyckoff position.    $\square_\mathrm{O}$ ( Wyckoff site $8e$)  was studied in the donor charge states 0, +1, +2. Several configurations with similar energies have been predicted for this defect \cite{Finazzi-2008}. Here we consider the split-vacancy configuration \cite{Morgan-2009} which should represent the ground state. We also considered the antisites O$_\mathrm{Ti}$ as donors in charge states 0, -1, -2, -3, -4 and the antisites Ti$_\mathrm{O}$ as acceptors in charge states 0, +1, +2, +3, +4. Antisites were scarcely studied in the literature, we only found them considered in the study of Ref. \onlinecite{Na-2006}. As their geometric structure has been barely considered before, we report it here for completeness. These defects originally occupy the same Wyckoff positions of the atoms they substitute. However, we found that for both antisite defects, the antisite atom relaxes along the $c$ direction from the ideal position of the atom it substitutes. Relaxation effects are particularly important for the Ti$_\mathrm{O}$ defect, due to the large radius of Ti atoms compared to the O one. Ti$_\mathrm{O}$ defects can be thought as a complex formed by a $\square_\mathrm{O}$ and a Ti$_i$. Such findings are summarized in Figure \ref{fig:antisites}.  Due to the large strain induced on the host material, such defects have a large formation energy and therefore a minor role in the defect chemistry of anatase.

The problem of localization of excess electrons in anatase has been discussed thoroughly in the literature due to the relevance of electron self-trapping for photocatalytic applications (see for example Ref. \onlinecite{Yin-2018} and references therein). It has been shows that the nature of these excess electrons is highly affected by the choice of the employed theoretical approach \cite{Spreafico-2014}. In particular, employing the the DFT+$U$ approach yields a different description of excess electron localization when different values of $U$ are employed \cite{Setvin-2014}. We calculated self-trapped electrons with the PBE+$U$[Ti] method and we found that it forms a small polaron with a formation energy (the polaron formation energy is defined as $E_{pol} = E_{loc}(N+1) - E_{deloc}(N+1)$, where $N$ is the number of electrons in the pristine system, $loc$ and $deloc$ denote electronic configuration in which the excess electron is localized on a Ti atom or delocalized in a conduction-band-like state, respectively) of around -0.3 eV, in agreement with the study of Setvin \emph{et al.} that found small polarons are stabilized in anatase when a $U$ value larger than 5 eV is used \cite{Setvin-2014}. We also found that the small polaron has a large formation energy, as given by equation \eqref{eq:formene},  of around 3.70 eV; therefore self-trapping appears unlikely  for any value of the electron chemical potential within the experimental band gap. Since experimental observations also suggests that excess electrons in anatase are not trapped unless other defects are present \cite{Setvin-2014}, we did not consider such species further.

The type of point defect studied with each computational approach is summarized in Table \ref{table:list}. For all approaches we considered the most important defects: Ti$_{i}$, $\square_\mathrm{O}$ and $\square_\mathrm{Ti}$ which are usually studied in the literature. The former two are expected to be the most important electron donors; while the latter the most important electron acceptor. Defects with higher formation energies, like O$_\mathrm{Ti}$ and Ti$_\mathrm{O}$ and the small polaron were  considered only within the PBE+$U$[Ti] approach.

\begin{table}
\caption{\label{table:list} Classes of point defects calculated with a given computational approach. $\times$ indicates that the given class of point defects was considered in all specified charge states. The charge states taken into account are: -2,-1,0,+1,+2 for  O$_{i}$, 0,+1,+2,+3,+4 for Ti$_{i}$, 0,+1,+2 for $\square_\mathrm{O}$, 0,-1,-2,-3,-4 for O$_\mathrm{Ti}$, 0,+1,+2,+3,+4 for Ti$_\mathrm{O}$ and -1 for the polaron}
\begin{ruledtabular}
 \begin{tabular}{ l c c c c } 
Defect &  PBEsol & PBE+$U$[Ti]  &  PBE+$U$[Ti,O] & HSE15  \\
\hline
O$_{i}$&  $\times$ & $\times$ & $\times$ & \\
Ti$_{i}$ & $\times$ & $\times$ & $\times$ & $\times$ \\
$\square_\mathrm{O}$ &  $\times$ & $\times$ & $\times$ & $\times$ \\
$\square_\mathrm{Ti}$ &  $\times$ & $\times$ & $\times$ & $\times$ \\
O$_\mathrm{Ti}$ &  & $\times$ & $\times$ &  \\
Ti$_\mathrm{O}$ &  & $\times$ & &  \\
polaron &  & $\times$ & 
\end{tabular}
\end{ruledtabular}
\end{table}

\subsection{\label{sec:TL} Transition levels} 
\begin{figure}
\includegraphics[height=1\textwidth]{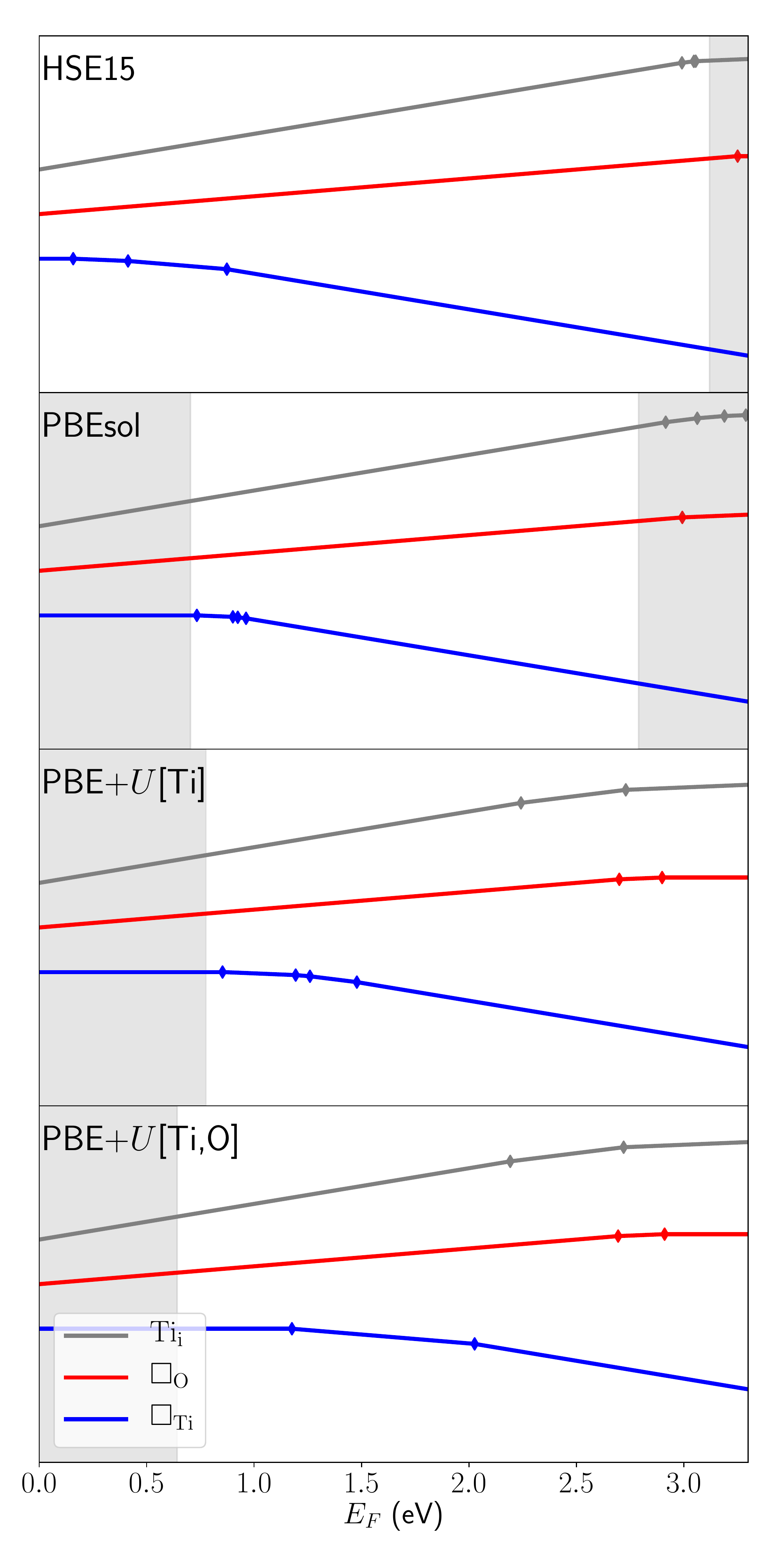}
\caption{\label{fig:ts} Graphical representation of the thermodynamic charge transition levels predicted by the four xc functionals employed in this study for Ti$_{i}$,  $\square_\mathrm{O}$ and $\square_\mathrm{Ti}$. The lines are obtained from the defect formation energies, whose values have been rigidly shifted for a more clear visualization of the transition levels. 
The zero of $E_F$ has been set at $\epsilon_V$ of HSE15 and valence band maxima have been aligned for all functionals, as explained in the text. Shaded areas show the band edges one would obtain without having aligned the valence band maxima.}
\end{figure}

As they are not affected by the chemical potentials of O and Ti, we start our discussion with the comparison of the thermodynamic charge transition levels, $\epsilon(q/q')$, among the different functionals. They are however affected by the predicted value of the valence band maximum, equation~\eqref{eq:translevel}, which must be aligned with respect to a common reference. This is illustrated in Figure~\ref{fig:ts} for the most relevant point defects in anatase: Ti$_{i}$,  $\square_\mathrm{O}$ and $\square_\mathrm{Ti}$, for which we calculated all the charge states for all four computational approaches. In Table~\ref{table:trans} we report the values of $\epsilon(q/q')$ predicted with the employed computational schemes, after having aligned the valence bands with the one predicted by HSE15 calculations.

 \begin{table}
\caption{\label{table:trans} Thermodynamic charge transition levels calculated after having aligned the valence band maximum to an unique reference (HSE15 top of the valence band). Values are in eV. A hyphen indicates that the charge transition level is predicted to not exist for the corresponding computational scheme. Transition levels appearing above the experimental band gap of 3.2 eV plus a small tolerance of 0.1 eV are excluded from the table except for comparing with another functional. In this case the values are shown in parenthesis.}
\begin{ruledtabular}
 \begin{tabular}{ l c c c c c} 

Defect & $q/q'$ & PBEsol & PBE+$U$[Ti]  &  PBE+$U$[Ti,O] & HSE15   \\ 
 \hline
\hline
\multirow{5}{4em}{Ti$_{i}$}  & 0/1 & 3.29 & (4.00) & (4.02) & (3.40)\\
                             & 1/3 & - & 2.73 & 2.72 & -\\
                             & 1/2 & 3.19 & - & - & 3.06\\
                             & 2/3 & 3.06 & - & - & 3.05\\
                             & 3/4 & 2.92 & 2.24 & 2.19  & 2.99\\
\hline
\multirow{3}{4em}{$\square_\mathrm{O}$} & 0/1 & 3.30 & 2.90 & 2.91 & - \\
                                        & 1/2 & 2.99 & 2.70 & 2.69 & -\\                                        
                                        & 0/2 & - & - & - & 3.25 \\
\hline
\multirow{6}{4em}{$\square_\mathrm{Ti}$} & -1/0 & 0.73 & 0.85 & - & 0.16\\
                                         & -2/0 & - & - & 1.18 & - \\
                                         & -2/-1 & 0.90 & 1.19 & - & 0.41\\
                                         & -3/-2 & 0.92 & 1.26 & - & -\\
                                         & -4/-3 & 0.96 & 1.48 & - & -\\
                                         & -4/-2 & - & - & 2.03 & 0.87\\
                                             
\end{tabular}
\end{ruledtabular}
\end{table}

One can notice that,  transition levels can have different qualitative behavior between functionals. 
The best agreement is obtained between PBE+$U$[Ti] and PBE+$U$[Ti,O], with a mean absolute error (MAE) of 0.14 eV. This should not be surprising as the inclusion of the same $U$ term on Ti atoms will give the same description of those defects where electrons are trapped on Ti sites, such as $\square_\mathrm{O}$. For   $\square_\mathrm{Ti}$ where PBE+$U$[Ti,O] yields larger electrons localized on the oxygen $2p$ orbitals, the agreement between between PBE+$U$[Ti] and PBE+$U$[Ti,O] is worse.
 One point that is noticeable in Table~\ref{table:trans} is that the agreement between HSE15 and PBEsol, once the VBM have been aligned, is rather good (MAE of 0.26 eV) so that standard semilocal functionals might be a better choice for a first estimation of the defect energetics, assuming the valence band maxima have been aligned with those of a more accurate functional.  This ability of semilocal functionals to give thermodynamic charge transition levels and defect formation energies in a rather good agreement with hybrid functionals is consistent with studies of other semiconductor materials, see for example references  \onlinecite{Alkauskas-2008,Lyons-2017}. It is somewhat surprising that the agreement between the PBE+$U$ approaches and HSE15 is the worst. One should have expected a better agreement as both PBE+$U$ and hybrid functionals partly correct for the self-interaction error which largely affects electron localization and thus the properties of the point defect. A more detailed analysis   for the understanding of the factors that affect the values of charge transitions levels and defect formation energies calculated with different functionals is in order. 

\begin{figure}
\subfloat[\label{fig:comp1vo}]{{\includegraphics[width=0.33\textwidth]{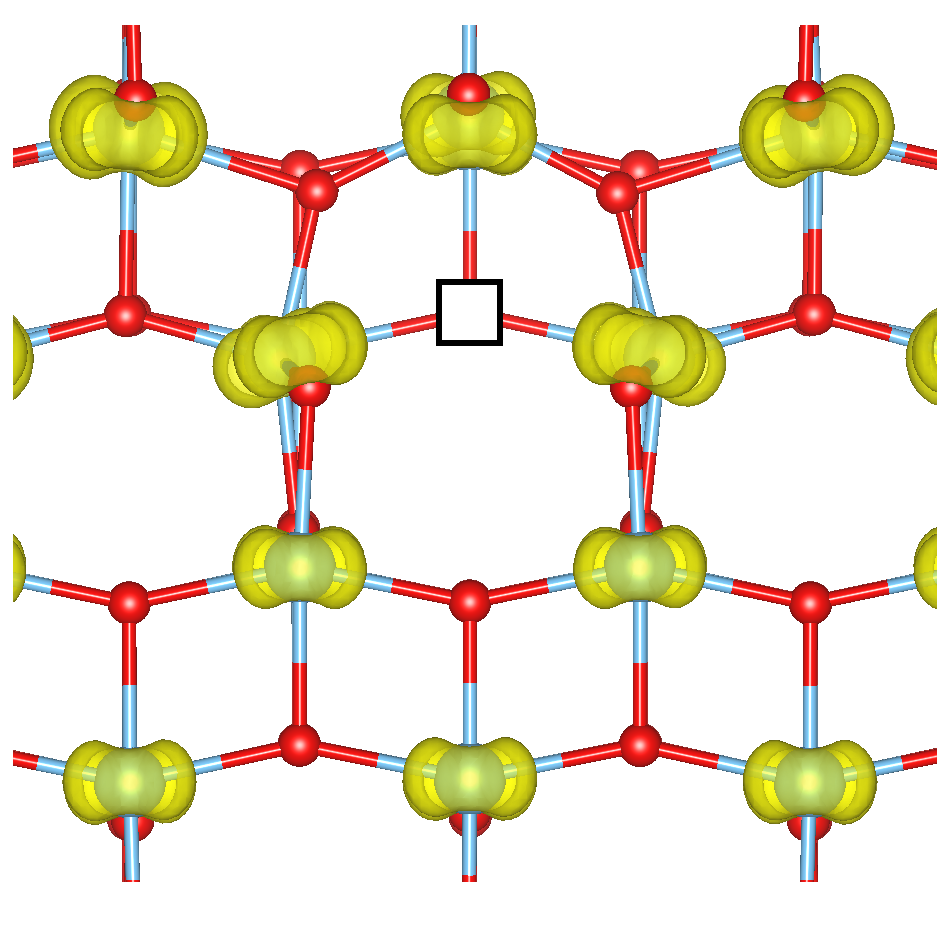} }} 
\subfloat[\label{fig:comp2vo}]{{\includegraphics[width=0.33\textwidth]{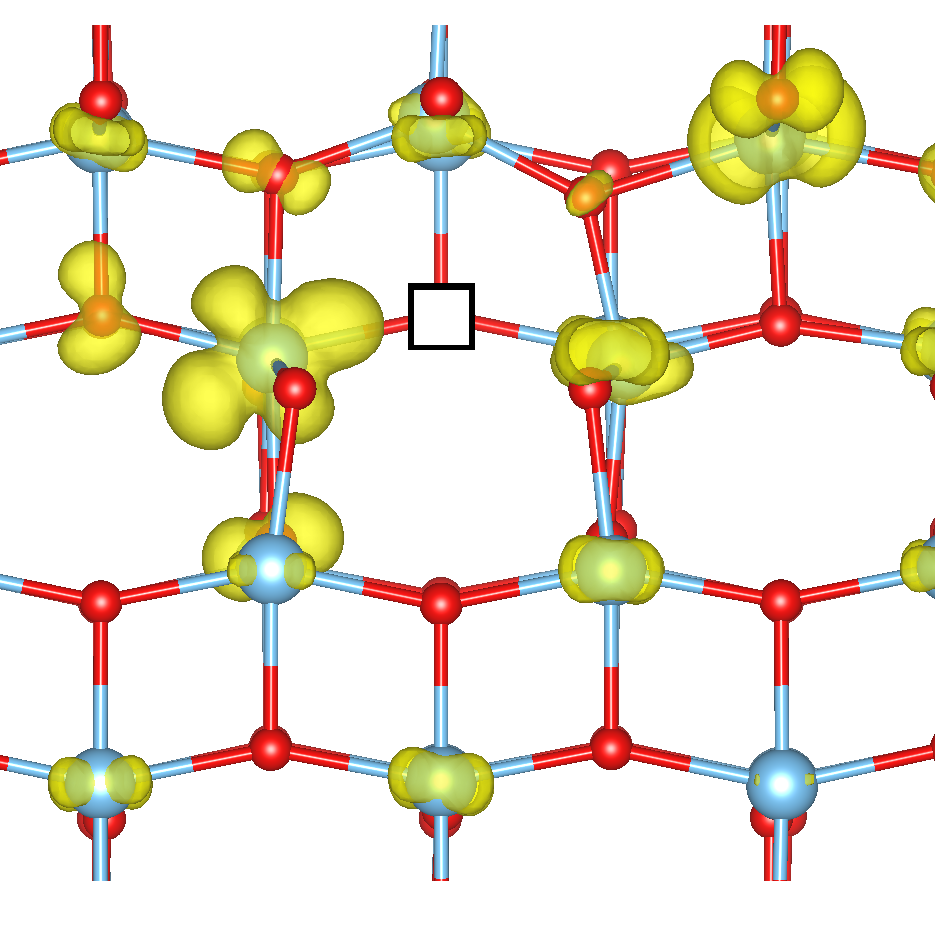}}} 
\subfloat[\label{fig:comp4vo}]{{\includegraphics[width=0.33\textwidth]{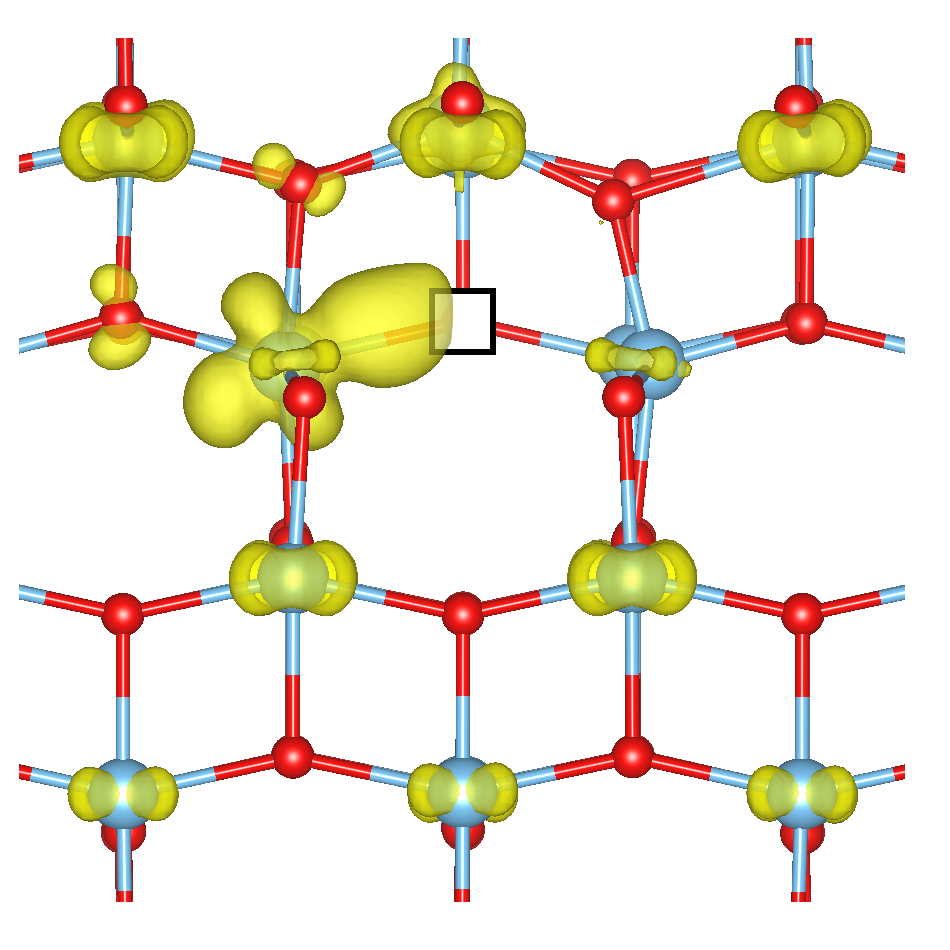}}}%
\caption{\label{fig:vo} Electronic charge density projected of the levels induced by the neutral  $\square_\mathrm{O}$ defect: (a) PBEsol, (b)  PBE+$U$[Ti],  
(c) HSE15.  Isosurfaces level  is shown at 0.005 e/\AA$^3$. The square represents the position of the oxygen vacancy. Oxygen atoms are shown in red. The plane of the figure is defined by the $\mathbf{b}$ and $\mathbf{c}$ cell vectors of the tetragonal cell.}
\end{figure}

 Taking the $\square_\mathrm{O}$, the main difference between PBEsol comes from the non-charged state being relatively more stable using HSE15, which results in a direct transition level from the +2 state to the noncharged state close the the CBM.
Figure \ref{fig:vo} shows a comparison of the geometric structure and defect-induced charge density of the neutral $\square_\mathrm{O}$ as obtained by the various xc functionals considered in this study. Removing an oxygen atom from pristine anatase reduces the point group symmetry to $mm2$. However, a more stable lower symmetry ($\bar{2}$ point group) configuration, named  a ``split-vacancy'' in an earlier GGA+$U$ study\cite{Morgan-2010}, can be found, Figure~\ref{fig:vo}. We were able to find these two configurations for all the employed functionals and found that the low-symmetry configuration has the lower energy than the high-symmetry one, even tough the difference is  sometimes small: 0.14 eV for PBEsol, 0.77 eV for PBE+$U$[Ti] 
and 0.38 eV for HSE15.   
As one can see from Figure \ref{fig:vo}, the symmetry breaking that characterizes the split-vacancy configuration is driven by electron localization on neighboring Ti atoms. Due to the self-interaction and band-gap errors, in PBEsol $\square_\mathrm{O}$ induces a defective level which is mixed with the bottom of the conduction band and the extra electrons are therefore delocalized on several Ti atoms. For this reason, the local distortions leading to the $\bar{2}$ symmetry are minimal. If we consider the PBE+$U$ approach instead, the electron localization is much stronger and affects mainly two distinct Ti atoms which become reduced to the +3 state. As a consequence, the local geometric distortions are also much more relevant, such features are found both with PBE+$U$[Ti] and PBE+$U$[Ti,O]. Also HSE15 predicts the defect state to be localized, albeit to a lesser degree than PBE+$U$. Taking HSE15 results as a reference, it is then evident that semilocal functionals tend to disfavor the localized solution due to self-interaction error. On the other hand, PBE+$U$ tends to overlocalize the electrons. For this approach, the ability to describe electron localization is governed by the $U$ term, and a value which gives a reliable description of the property of the host material (such as band gap and cell parameters) is not necessarily suitable for describing the defective system. We mention that previous HSE studies\cite{Deak-2015,Boonchun-2016}  only considered the high-symmetry configuration which would lead to an overestimation of the energy of the non-charged $\square_\mathrm{O}$ defect. 

These results underline an important advantage of the PBE+$U$ approach: even if the total energies have limited predictive value when comparing different oxidation states, PBE+$U$ allows a computationally relatively efficient exploration of the potential energy surface where also structures characterized by localized states are present. These can then serve as a starting points for more time consuming methodologies like the use of hybrid functionals.

\begin{figure}
\subfloat[\label{fig:comp1v}]{{\includegraphics[width=0.33\textwidth]{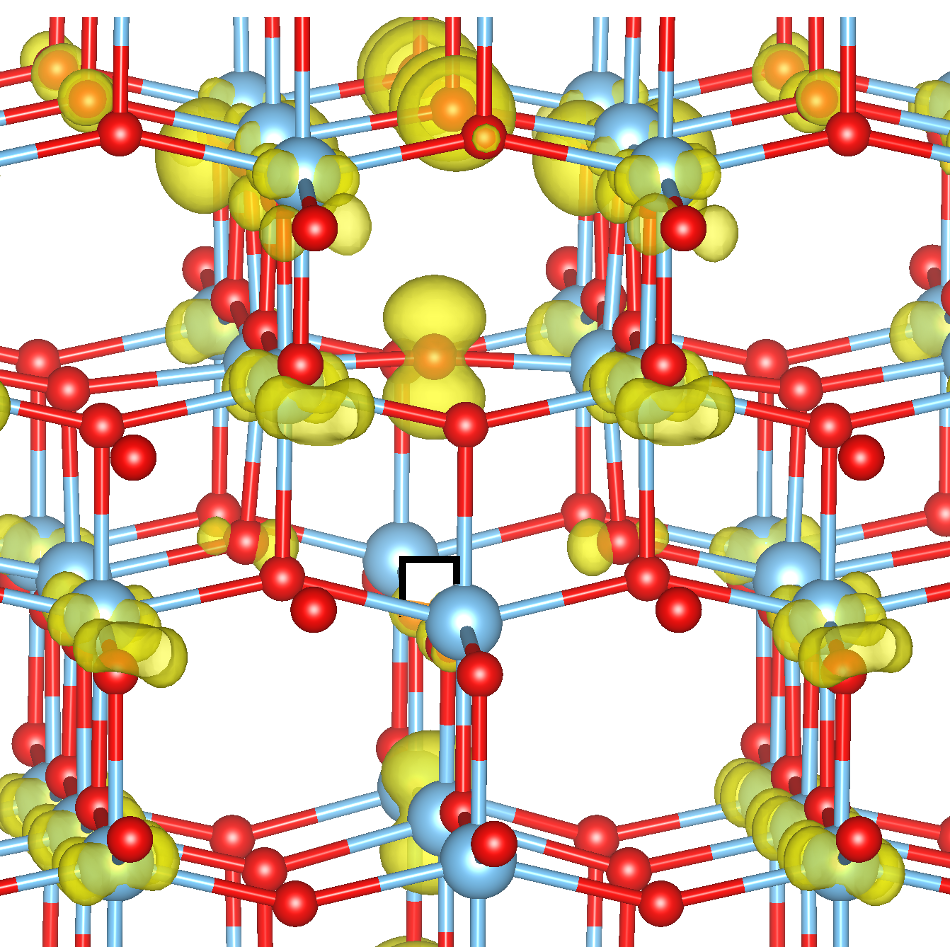} }} 
\subfloat[\label{fig:comp2v}]{{\includegraphics[width=0.33\textwidth]{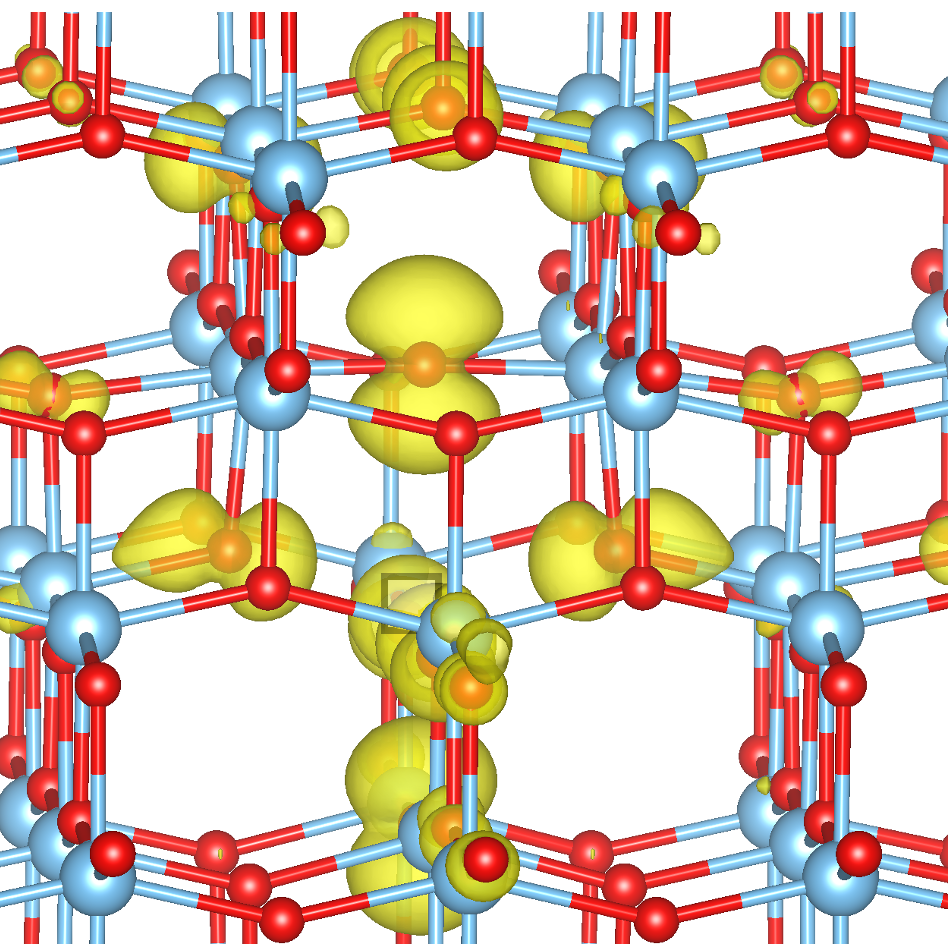}}} 
\subfloat[\label{fig:comp4v}]{{\includegraphics[width=0.33\textwidth]{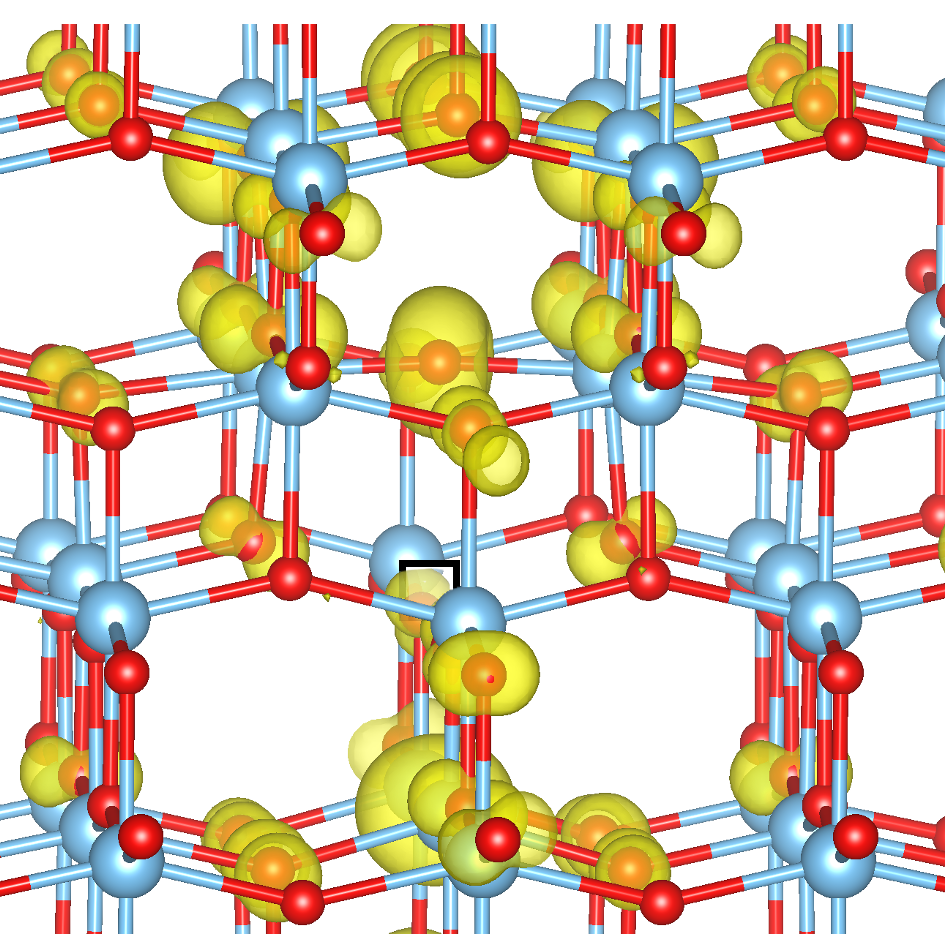}}}%
\caption{\label{fig:vti} Electronic charge density projected on the defective levels induced by the fully ionized (-4) $\square_\mathrm{Ti}$ defect : (a) PBEsol, (b)  PBE+$U$[Ti], (c) HSE15.  Isosurfaces level  is shown at 0.005 e/\AA$^3$. The square represents the position of the titanium vacancy. Oxygen atoms are shown in red. The plane of the figure is defined by the $\mathbf{b}$ and $\mathbf{c}$ cell vectors of the conventional tetragonal cell.}
\end{figure}

Analogous considerations hold for the other studied defects. In particular, we show in Figure \ref{fig:vti} the comparison for the most relevant electron acceptor: the $\square_\mathrm{Ti}$. We choose the completely ionized (4-) defect state in order to emphasize the behaviour of localized electrons. In this charge state, four electrons have been accepted from the host material and the defect-induced levels are filled.
 From Figure \ref{fig:vti} one can see that  the PBE+$U$ approach  predicts electron localization mostly on four oxygen atoms neighboring the vacancy: two oxygen atoms are the nearest neighbors located on the line parallel to the [001] direction and other two oxygen atoms are the closest neighbors along the line parallel to the [010] direction. 
 Electron localization on these sites is similarly predicted by the PBE+$U$[Ti] and PBE+$U$[Ti, O] schemes, but in the latter more electrons are localized on the former two oxygen atoms than the latter two.
 Once more, the extra charge is more delocalized in PBEsol. The HSE15 predicts a rather localized electronic configuration, especially on the two oxygen atoms along the [001] direction, but to a noticeably lesser degree than in PBE+$U$. 

\subsection{\label{sec:res_chempots} Bulk Thermochemistry}
Calculated values for the formation enthalpies of selected titanium oxides are reported in Table \ref{table:thermo}. We took those compounds whose crystal structures and experimental formation enthalpies are well described on the NIST-JANAF thermochemical tables \cite{NIST}. In particular, we took: TiO$_2$ anatase, TiO$_2$ rutile, Ti$_2$O$_3$, Ti$_3$O$_5$ and TiO.
 TiO$_2$ rutile has a tetragonal primitive cell (space group $P4_2/mnm$) and it is a semiconductor with a band gap of 3.0 eV \cite{Coronado-2008}. For Ti$_2$O$_3$, we took the rhombohedral phase (space group $R\bar{3}c$) which is commonly found at room temperature. Its ground state is non-magnetic and the system is a semiconductor with a very small band gap of around 0.1 eV, given by a trigonal crystal field distortion that splits the occupied Ti $3d \, a_{1g}$ states from the unoccupied $3d \,  e_g$ ones \cite{Guo-2012}. The computational modeling of this phase is challenging. PBEsol predicts a metal while with PBE+$U$[Ti] we obtain a very large band gap of around 1 eV. Moreover, PBE+$U$[Ti] finds that the antiferromagnetic state has a much lower energy than the diamagnetic one. Due to these problems,  we excluded such compound from the fitting procedure described by Jain \emph{et al.} \cite{Jain-2011}. Ti$_3$O$_5$ was taken in the low-symmetry phase which is stable for temperatures below 120$^\circ$C and has the monoclinic structure (space group $C2/m$)\cite{Asbrink-1959}. It is a semiconductor with a small gap of 0.14 eV, which also arises from the splitting of the Ti $3d$ states \cite{Ohkoshi-2010}. While semi-local functionals predict it to be a metal, the DFT+$U$ approach correctly describes the semiconductor state \cite{Liu-2014}. Finally, TiO has the NaCl structure, with a cubic cell and space group $Fm\bar{3}m$. This compound is metallic \cite{Denker-1966} and was therefore  not calculated using the PBE+$U$ setup.

As shown by Table~\ref{table:thermo}, both PBEsol and PBE+$U$ predict the ground state of the TiO$_2$ system to be anatase. We found that anatase is also more stable than rutile using HSE15, which predicts anatase to be more stable than rutile by around 93 meV per formula unit. 
Regarding the other phases, PBEsol gives formation enthalpies in good agreement with the experimental data. That this is the case, despite the varying oxidation states, is somewhat surprising, as the self-interaction error should favor states with a smaller number of localized $d$ electrons, \emph{i.e.} more oxidized states. However, GGA functionals tend also to overbind the O$_2$ molecule \cite{Wang-2006}. This overbinding will on the other hand favor more reduced states, and the good agreement observed for PBEsol can to a certain degree be attributed to this cancellation of errors. 

In the employed DFT+$U$ formulation\cite{ldaUfll,Dudarev-1998}, an energy contribution of the form:
\begin{equation}\sum_\sigma
 \frac{U}{2} \mathrm{Tr} \left[\mathbf{n}_\sigma(1 - \mathbf{n}_\sigma)\right]
\end{equation}
is added to the energy functional. $U$ quantifies an  effective on-site coulombic repulsion that is applied to chosen atomic sites.  $\sigma$ is the spin variable and $\mathbf{n}$ the occupation number matrix in the localized orbital representation.  The energy contribution is positive for all fractional occupation numbers and it so increases the energy of Ti-O systems, where the $d$-orbitals are not fully localized. This results in a large positive energy contribution to all systems where the $U$ is applied. In the present case, where we would not apply the $U$ for the Ti metal, this would lead to a strong underestimation of the energy gain by forming the oxide, as shown in the column  ``PBE+$U$[Ti]'' of Table~\ref{table:thermo}, which reports the calculated formation enthalpies where the Hubbard $U$ is used for the non-metallic oxides of titanium. It is then clear that formation enthalpies obtaining by mixing PBE and PBE+$U$ calculations do not have any physical meaning.  To improve on this, Jain \emph{et al.} proposed a method that adjusts the energies calculated with GGA+$U$ in such a way that they can be mixed with pure GGA calculations \cite{Jain-2011}. Their method is based on fitting formation energies of binary transition metal oxides in order to find a correction term per transition metal atom that has to be added to the calculated GGA+$U$ energies \cite{Jain-2011}. The value of the correction term we found from this fitting is of $\Delta E_\text{Ti} = 2.89$~eV per Ti atom. The results obtained after adding this correction together with the correction term for the GGA overbinding\cite{Wang-2006} are summarized in the "Corrected" column ``PBE+$U$[Ti] Corrected''. Of course, applying such correction will drastically improve the calculated formation enthalpies for the compounds included in the fitting: the MAE drops to 14~meV/atom.

\begin{figure}
\includegraphics[width=1\textwidth]{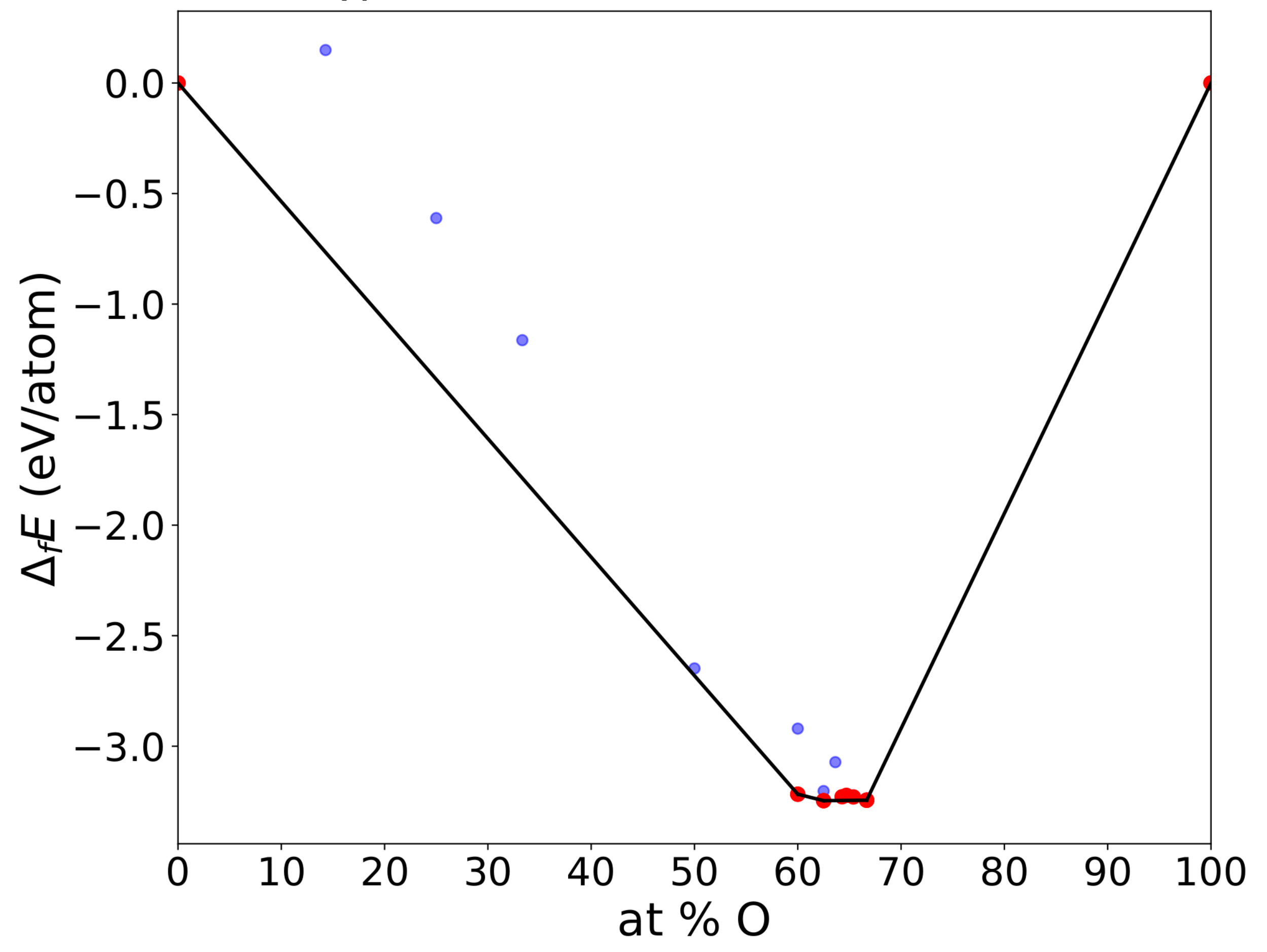}
\caption{\label{fig:chull} Convex hull calculated in the corrected PBE+$U$[Ti] setup. The red dots represent compounds that lie on the hull or have a distance form it within 25meV/atom. Note that the Ti$_2$O$_3$ system on the hull has the antiferromagnetic state; while the non-magnetic one has higher formation energy. }
\end{figure}

\begin{table}
\caption{\label{table:thermo} Formation enthalpies of selected oxides of titanium calculated with different computational setups. 
 In the ``PBEsol'' column values are obtained by calculating the electronic energies of all species using the PBEsol functional. In the ``PBE+$U$[Ti]'' column, values are obtained by calculating the electronic energy with the PBE+$U$[Ti] setup for all compounds except TiO, Ti and O$_2$, for which PBE was used.  In the ``PBE+$U$[Ti] Corrected'' column we apply the correction for the elements as detailed in the main text. All values are reported in eV per formula unit. Note that for the PBE+$U$ scheme the table reports the antiferromagnetic phase of Ti$_2$O$_3$, which is predicted to be the ground-state by this framework. Experimental values represent the formation enthalpies at room temperature and standard pressure. Computed values report the ground-state zero-temperature and zero-pressure values. The last row  reports the calculated mean absolute error from the experimental data.}
\begin{ruledtabular}
\small
\begin{tabular}{lrrrrr}
Comp. & PBEsol & PBE+$U$[Ti] &  PBE+$U$[Ti] &  Exp.\footnote{Ref. \onlinecite{NIST}} \\
& & & Corrected & \\
\hline
Anatase &  -9.53 & -5.46 & -9.73&  -9.73 \\
Rutile & -9.47 &-5.45  & -9.73 &  -9.79\\
Ti$_2$O$_3$ & -15.26 & -8.21 & -16.09 &-15.76 \\
Ti$_3$O$_5$ & -24.89  & -13.47 & -25.62 & -25.49 \\
TiO & -5.39  &-4.62 & -5.30 & -5.58 \\
MAE & 0.36 & 6.13 &  0.14   & 0 \\
\end{tabular}
\end{ruledtabular}
\end{table}

An accurate determination of the thermochemistry of the relevant phases is important in the calculation of defect formation energies as it determines the limit values of the chemical potentials of Ti and O.
From the experimental formation enthalpies it results that the first phase that starts to precipitate in Ti-rich conditions is Ti$_2$O$_3$. In order to obtain the competing phases in such a limiting conditions as predicted by the various xc functionals we calculated the convex hull of the Ti-O systems. Figure \ref{fig:chull} shows shows the hull calculated using the PBE+$U$[Ti] setup with Jain \emph{et al.}'s correction scheme (PBE+$U$[Ti] Corrected). In order to calculate the hull, we started from the compounds and energies available on the AFLOW repositories \cite{Curtarolo-2012,Taylor-2014}, we performed a first screening and selected all the compounds lying within 0.25 eV/atom from the convex hull. For such compounds, we then calculated the total energies using our PBE+$U$[Ti] scheme and recalculated the convex hull using the new energies corrected  with Jain \emph{et al.}'s method. From these calculations, we found that the first phase that precipitates in Ti-rich conditions is not the non-magnetic phase of Ti$_2$O$_3$ but the anti-ferromagnetic one, which lies on the convex hull in Figure \ref{fig:chull}. The non-magnetic phase of Ti$_2$O$_3$ has a much larger energy and considering it as the ground state phase would make Ti$_3$O$_5$ to be the first compound to segregate in Ti-rich conditions.
 This shows that care must be taken when one attempts to calculate the value of  the chemical potentials of Ti and O in the Ti-rich and O-rich limits from functionals which do not correctly describe the thermochemistry of the system. 
 
 Table \ref{table:mus} reports the values of the chemical potentials, with respect to their standard state, calculated using different methods for the O-rich and Ti-rich limits and considering only phase competition between anatase, the sesquioxide (in the ground state electronic structure predicted by that functional) and pure Ti and O$_2$. We can notice that the ranges of $\mu_\text{O}$ and $\mu_\text{Ti}$ for O-rich conditions agree well among PBEsol, PBE+$U$[Ti], HSE06 (obtained from reference \cite{Boonchun-2016}) and the experimental data. For PBE+$U$[Ti] this happens since both the energy of the O$_2$ molecule and of anatase are fitted to experimental data, as described above. While, as we mentioned before, PBEsol is expected to be affected both by the GGA shortcomings in the binding energy of O$_2$ and by the self-interaction error, which tend to compensate in pure GGA functionals and a good agreement with the experiment is found.

If Ti-rich conditions are considered, now the relevant phases for determining the chemical potential ranges, as found from experimental data, are TiO$_2$ anatase and Ti$_2$O$_3$. While HSE06 does a fairly good job in predicting the thermochemistry of transition-metal compounds, as already mentioned, compensation of errors are present in PBEsol, which also gives values close to the experimental ones. Also the corrected PBE+$U$[Ti] results tend to agree with the experimental ones; but while both PBEsol and HSE15 predict a value of  $\Delta \mu_\text{Ti}$ larger than the one derived from experimental data, the corrected PBE+$U$[Ti] predicts smaller values. Such discrepancy would add another source of disagreement between functionals on the calculated defect formation energies in the Ti-rich limit.

 \begin{table}
\caption{\label{table:mus} Chemical potential values for Ti and O calculated with different methods when O$_2$ and Ti$_2$O$_3$ are the competing phases in O-rich and Ti-rich conditions, respectively. Values in the Exp. column are obtained from the experimental formation enthalpies reported in Table \ref{table:thermo}.}
\begin{ruledtabular}
 \begin{tabular}{p{0.6 cm} l c c c c} 

 $\Delta \mu$   & (eV) & PBEsol & PBE+$U$[Ti]    & HSE06\footnote{Ref. \onlinecite{Boonchun-2016}} &  Exp.   \\ 
  & & & Corrected & \\
 \hline
\multirow{2}{4em}{$\Delta \mu_\text{Ti}$}  & O-rich   & -9.53 &  -9.73 & -9.76  & -9.73 \\
  							     & Ti-rich & -1.92   & -2.97& -1.72    &  -2.34  \\ 
\multirow{2}{4em}{$\Delta \mu_\text{O}$}  & O-rich  & 0 & 0 &  0 & 0  \\
  							     & Ti-rich &  -3.80  &  -3.38 & -4.02     & -3.70
\end{tabular}
\end{ruledtabular}
\end{table}

\subsection{\label{subsec:comparison} Defect Formation Energies}

\begin{figure}
\subfloat[\label{fig:comp1}]{{\includegraphics[width=0.5\textwidth]{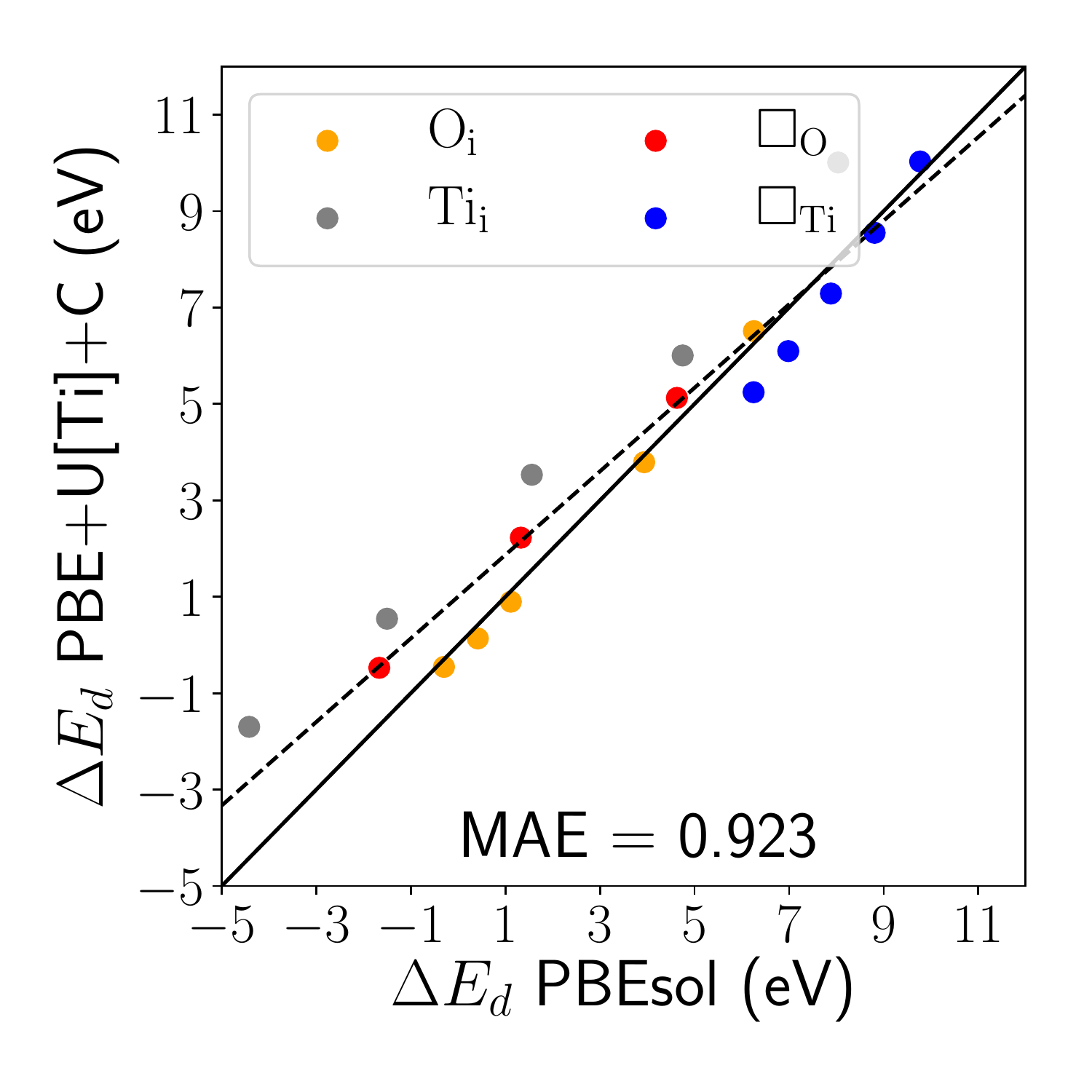} }} 
\subfloat[\label{fig:comp2}]{{\includegraphics[width=0.5\textwidth]{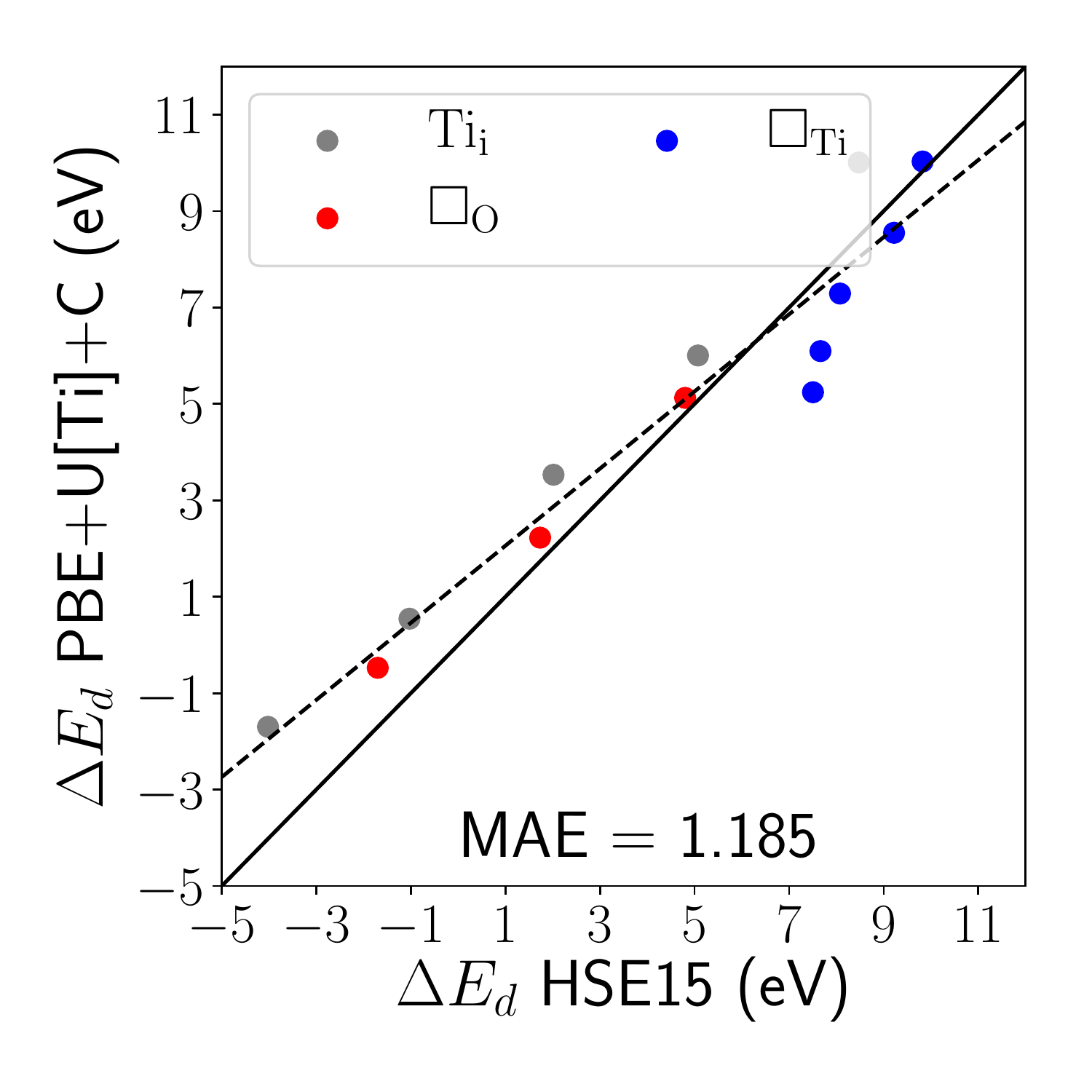}}} \\
\subfloat[\label{fig:comp3}]{{\includegraphics[width=0.5\textwidth]{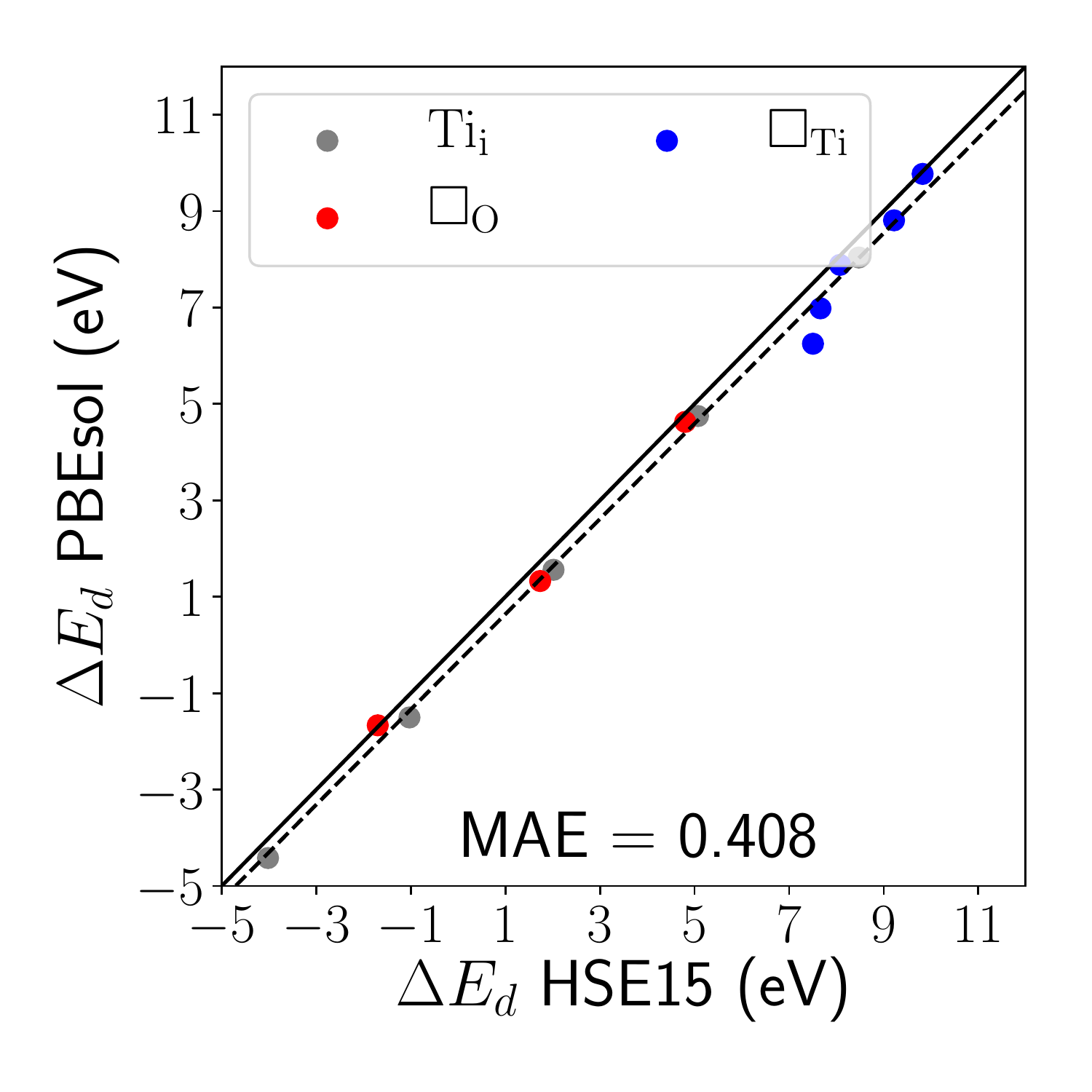}}} 
\subfloat[\label{fig:comp4}]{{\includegraphics[width=0.5\textwidth]{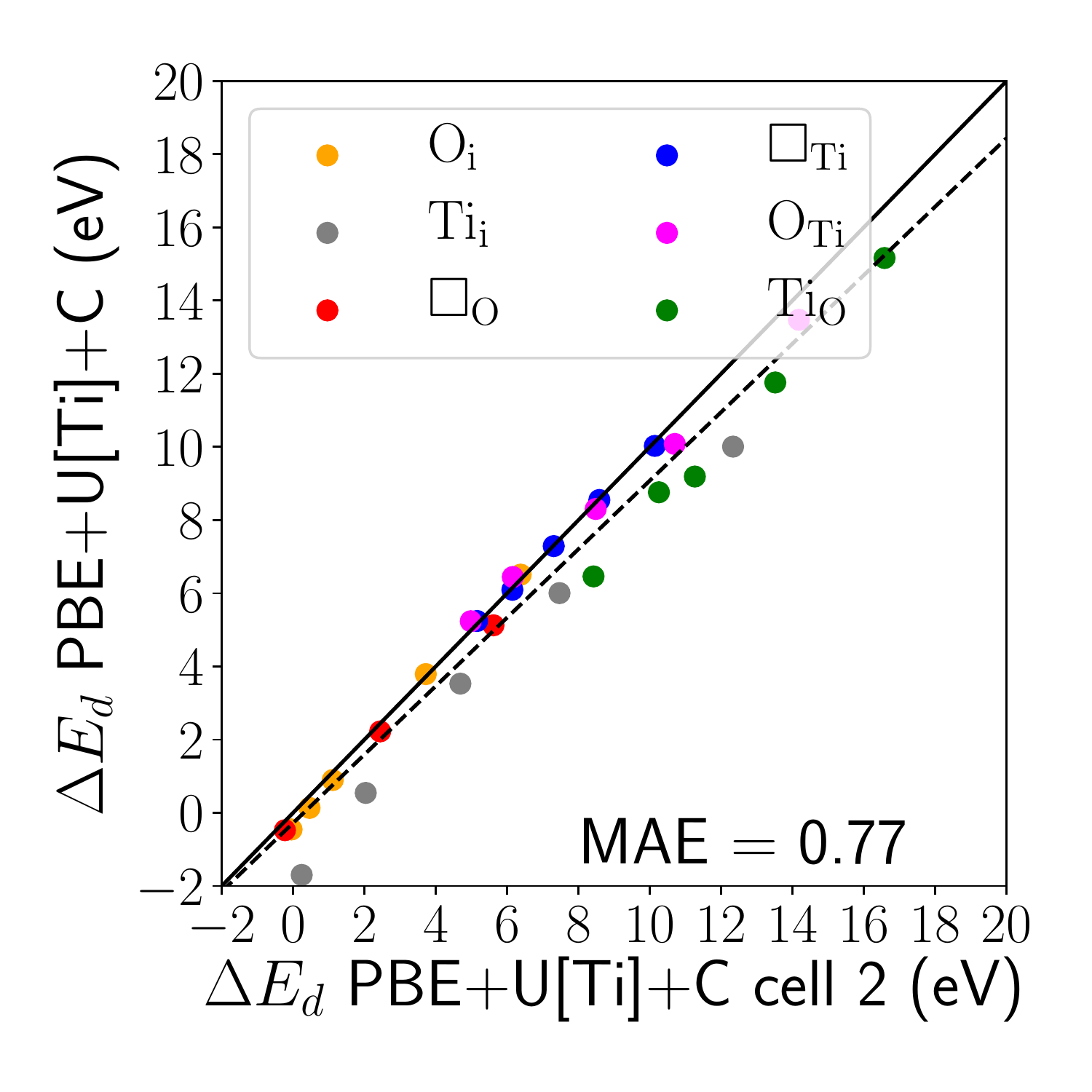}}}%
\caption{ Comparison between the formation energies of point defects in different charge states for O-rich conditions between: (a) PBEsol and corrected PBE+$U$[Ti] (PBE+$U$[Ti]+C), (b) HSE15 and PBE+$U$[Ti]+C, (c) HSE15 and PBEsol, (d) PBE+$U$[Ti]+C and PBE+$U$[Ti]+C where the cell parameter has been fixed to the PBEsol value (cell 2).  The bold black line indicates the ideal case where both schemes give the same formation energies. The dashed line is obtained from linear regression. }
\end{figure}

We calculated defect formation energies in the O-rich limit, where the chemical potential of the oxygen is set equal to half the electronic energy predicted for O$_2$ and the relevant thermochemistry is in good agreement among all xc functionals and experiments as well, and compare them among the different employed xc functionals, as shown in Figure \ref{fig:comp1}. 
The analysis performed in Section \ref{sec:res_chempots} should make obvious that formation energies obtained by mixing GGA+$U$ and GGA are unreliable without applying some correction scheme. We thus  compared  only the formation energies calculated with PBEsol, the corrected PBE+$U$[Ti] and HSE15 after the band alignment procedure. 

One can notice that the agreement between calculated formation energies is quite poor and that the MAEs are quite large: 0.92 eV between PBE+$U$[Ti] and PBEsol, 1.2 eV between the corrected PBE+$U$[Ti] and HSE15. the best agreement is obtained between PBEsol and HSE15, with a MAE of 0.41~eV which is comparable to the one found between PBE and HSE06 for GaN defects \cite{Freysoldt-2016}.  
One also notices that while PBEsol gives almost exclusively defect formation energies which are smaller than those predicted by HSE15, the formation energies predicted by the corrected PBE+$U$[Ti] do not follow a general trend. This also partly explains why the transition levels predicted with this functional are in such a poor agreement with HSE15, as transition levels are given by differences in defect formation energies. 

To end this section,  we assess the effect of fixing cell parameters to experimental values or values that do not correspond to the ground state predicted by  given functional for anatase. For this purpose we used the corrected PBE+$U$[Ti] scheme in order to calculate the formation energies of point defects in two cases: in the first case, the cell parameter was fixed to the value predicted  by the PBE+$U$[Ti] method. We call this choice of cell ``cell 1''. In the second case, we instead fixed the cell parameter to the value predicted by PBEsol, which is in better agreement with the experimental one. We refer to this second choice of cell parameter as ``cell 2''. PBE+$U$[Ti] overestimates the cell parameter of anatase; this entails that using ``cell 2'' induces a large external pressure on the supercell, from the diagonal part of the stress tensor we found that such pressure is very large, around 16 GPa.  Figure \ref{fig:comp4} summarizes the differences in formation energies predicted using the two different cells. As expected, using ``cell 2'' overestimates the defect formation energy, since an external pressure is applied on the supercell. While the MAE is quite large, around 0.77 eV, we notice that for the largest part, the discrepancy can be assigned to the Ti$_i$ and Ti$_\mathrm{O}$ classes of defects. This is not surprising as we mentioned before that both defects exert a large strain on the host material since they are characterized by the presence of an atom such as Ti, with a large atomic radius, occupying an interstitial site. Therefore, fixing cell parameters to values other than those predicted by the used theoretical scheme, can lead to considerable errors if the considered defects induce large strains on the crystal and is not a recommendable practice.

\section{Conclusions} 
In summary, we have found several factors that can affect the values of defect formation energies and charge transition levels in anatase and make comparisons between different functionals difficult.

Considering charge transition levels and defect formation energies, we find that, after band edges have been properly aligned to a common reference, the best agreement between functionals are found to occur between the PBEsol approaches and HSE15. This fact appears counter intuitive as PBE+$U$ is able to give defect geometric and electronic structures  in closer agreement to HSE15 than PBEsol, by partly correcting for self-interaction errors.  Also when defect formation energies are considered, there is a fair agreement between PBEsol and HSE15 values.

These observations suggest that although semilocal functionals are inadequate to correctly describe the geometric and electronic properties of defect where charge-localization is relevant, they might be a good choice for a first estimation of the energetic properties of point defects in semiconductors, assuming the valence band maximum is correctly aligned using a more accurate functional. On the other hand, the use of DFT+$U$ is advantageous for obtaining a more accurate initial geometric and electronic configuration which can be used as a starting point for more accurate and computationally expensive functionals.

\section{Acknowledgments}
The authors acknowledge support from the Austrian Science Funds (FWF)  under project CODIS (FWF-I-3576-N36). We also thank the Vienna Scientific Cluster for providing the computational facilities (1523306: CODIS).

\nocite{*}
\bibliographystyle{unsrtnat}
\bibliography{references}

\end{document}